\documentstyle[twoside,fleqn,npb,epsfig,amsmath,here]{article}

\def \ps {p\hspace{-0.43em}/}
\def \rt {r_t}
\def \ru {r_u}
\def \bk {\nonumber \\ && \hspace{-1.5cm}}
\def \bkd {\nonumber \\ && \hspace{-2cm}}
\def \bkr {\right. \nonumber \\&& \left. \hspace{-1.5cm}}

\title{Massive two-loop Bhabha scattering -- the factorizable subset
\thanks{Talk given by A.W. at RADCOR and Loops and Legs 2002 in Banz, Germany, to appear in the proceedings.}
      }

\author{J. Fleischer\address{Fakult\"at f. Physik, Universit\"at Bielefeld,   Universit\"atsstr. 25,  33615 Bielefeld, Germany}
        \thanks{Work supported in part by the EC under contract HPRN-CT-2000-00149 Physics at Colliders and by the EC under contract MCIF-2001-01652.
}
, T. Riemann\address{Deutsches Elektronen-Synchrotron, DESY Zeuthen, Platanenallee
  6, 15738 Zeuthen, Germany },
        O.~V. Tarasov$^{a, b}$
    and A. Werthenbach$^{b, }$\address{Theory Division, CERN, 1211 Geneva 23, Switzerland}}

\begin{document}

\allowdisplaybreaks

\begin{abstract}
The experimental precision that will be reached at the next generation of colliders makes it indispensable to improve theoretical predictions significantly. Bhabha scattering $( e^+ \, e^- \to e^+\, e^-)$ is one of the prime processes calling for a better theoretical precision, in particular for non-zero electron masses. We present a first subset of the full two-loop calculation, namely the factorizable subset.  Our calculation is based on DIANA \cite{Tentyukov:1999is}. We reduce tensor integrals to scalar integrals in shifted (increased) dimensions and additional powers of various propagators, so-called dots-on-lines. Recurrence relations \cite{Tarasov:1996br,Fleischer:1999hq} remove those dots-on-lines as well as genuine dots-on-lines (originating from mass renormalization) and reduce the dimension of the integrals to the generic $ d = 4 - 2 \epsilon $ dimensions. The resulting master integrals have to be expanded to ${\it O }(\epsilon )$ to ensure proper treatment of all finite terms. 
\end{abstract}
\maketitle
\vspace*{-11cm}
\noindent
BI-TP 2002/27 \\
CERN 2002-306  \\
DESY 02-196  \\
hep-ph/0211167
\vspace*{8.2cm}
\setcounter{footnote}{0}
\vspace{-0.3mm}
\section{Motivation}
At the next generation of linear $e^+ \, e^-$ colliders the experimental precision of $ {\it O}(1/1000)$
will exceed current standards by one  order of magnitude. To ensure the success and strengthen the physics case of such a machine, a precise understanding of the theoretical predictions is
of paramount importance. In particular Bhabha scattering is important  for luminosity monitoring, and hence a vital ingredient of almost every measurement. Recent studies \cite{Bern:2000ie,Glover:2001ev} have focused on the calculation of the two-loop QED matrix element to massless Bhabha scattering, i.e. the photon and the electron are treated as massless particles. To further increase the theoretical understanding of the Bhabha cross section and to control collinear singularities  we consider the electron to be a massive particle.
At the energy scale foreseen for the next generation of $e^+ \, e^-$ colliders, electroweak corrections cannot be ignored. While it is certainly a good approximation to consider small angle Bhabha scattering to be highly dominated by pure QED corrections, this does not apply to
the whole phase space. Taking the electron mass into account in the genuine QED calculation will open the door to more involved multi-scale electroweak higher order calculations. Here we are presenting the first progress made towards a complete two-loop calculation. 
\section{One-loop contributions}
At the Born level the Bhabha scattering process $e^-(p_1) \, e^+(p_4) \to \gamma^* \to e^-(-p_2) \, e^+(-p_3)$ is characterized through an $s$- and a $t$-channel photon exchange diagram.
One-loop radiative corrections constitute the two-loop calculation in the form of $(1 \, {\rm loop} )^2$ contributions to the
cross section. To generate the diagrams in the form of a Postscript file, as well as in symbolic notation for further algebraic manipulation, we use DIANA, the DIagram ANAlyser  by Tentyukov and  Fleischer \cite{Tentyukov:1999is}~\footnote{For some very recent valuable features of DIANA see \cite{Tentyukov:2002ig} as well as \cite{Fleischer} in these proceedings.}.  We choose to work in the Feynman gauge~\footnote{In fact, part of the calculation has been performed in a general $R_{\chi}$ gauge.}. This has the advantage that tadpole contributions are energy independent and hence do not contribute to the physical cross section. Self-energy, vertex and box diagrams make their appearance in equal parts as  $s$- and  $t$-channel  diagrams. The following ten diagrams have been calculated.\\ 
\vspace*{-1cm}
\begin{figure}[H]
  \begin{center}
    \epsfysize=11cm
    \epsffile{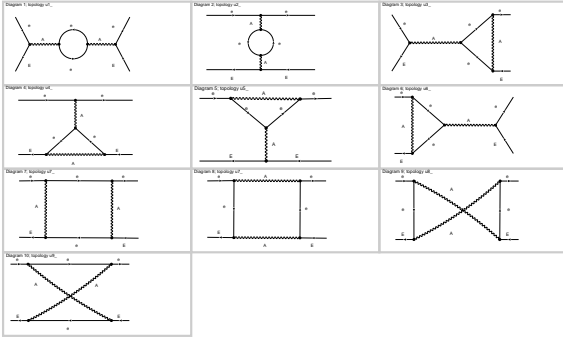}
   \vspace{-8.5cm}
    \caption{One-loop diagrams to the process $e^+ e^- \to e^+ e^-$.}
    \label{fig:1loop}
  \end{center}
\end{figure}
\vspace*{-1cm}
\noindent
We decompose the full one-loop matrix element into 12 amplitudes, that is in fact into 24 amplitudes, counting the $s$- and  $t$-channel contributions separately. For compactness we omit the spinors. The richness in structure is indeed only required for the box contributions. The self-energies and vertices contribute to only two of the structures below. We would like to remark at this point that the products of Dirac                                      $ \gamma$ matrices ${\cal M}_9$ to $ {\cal M}_{12}$ can in principle be simplified at the cost of artificially introducing $\gamma_5 $, which we decided to avoid altogether. 
\begin{align}
    {\cal M}_1 & =  \left[  \, 1   \,  \times   \,  1  \,  \right]   \, F_1    \nonumber \\
   {\cal M}_2 & =  \left[ \, \ps_4    \,  \times   \,  1   \,  \right]   \, F_2    \nonumber \\
  {\cal M}_3 & =  \left[  \, 1    \,  \times   \,  \ps_2   \,   \right]   \, F_3    \nonumber \\
 {\cal M}_4 & =  \left[  \, \ps_4    \,  \times   \,  \ps_2   \,   \right]   \, F_4    \nonumber  \\
 {\cal M}_5 & =  \left[ \,   \gamma_{\mu}   \,  \times   \,   \gamma_{\mu}    \,   \right]   \, F_5    \nonumber \\
  {\cal M}_6 & =  \left[ \, \gamma_{\mu}  \, \ps_4     \,  \times   \,   \gamma_{\mu}    \,   \right]   \, F_6    \nonumber \\
  {\cal M}_7 & =  \left[ \, \gamma_{\mu}       \,  \times   \,
\gamma_{\mu} \, \ps_2    \,  \right]   \, F_7    \nonumber \\
  {\cal M}_8 & =  \left[ \, \gamma_{\mu}  \, \gamma_{\nu}      \,
\times   \,  \gamma_{\nu}\,   \gamma_{\mu}     \,  \right]   \, F_8    \nonumber \\
  {\cal M}_9 & =  \left[ \, \gamma_{\mu}  \,  \gamma_{\nu} \, \ps_4     \,  \times   \,  \gamma_{\nu} \,  \gamma_{\mu}     \,   \right]   \, F_9    \nonumber \\
  {\cal M}_{10} & =  \left[ \, \gamma_{\mu}  \,  \gamma_{\nu}     \,
\times  \,  \gamma_{\nu}  \, \gamma_{\mu}  \, \ps_2    \,   \right]   \, F_{10}    \nonumber \\
  {\cal M}_{11} & =  \left[ \, \gamma_{\mu}   \,  \gamma_{\nu} \, \ps_4
\,  \times  \,  \gamma_{\nu}  \,  \gamma_{\mu}  \, \ps_2  \,  \right]   \, F_{11}    \nonumber \\
  {\cal M}_{12} & =  \left[ \,  \gamma_{\mu}   \,  \gamma_{\nu}  \,  \gamma_{\rho}  \,  \times \,  \gamma_{\rho}  \,  \gamma_{\nu}  \,  \gamma_{\mu}     \,  \right]   \, F_{12}    
\end{align}
The entire  calculation is parametrized  through the corresponding 12 form factors for every diagram. 
Symmetries allow to further compress the calculation. This is in particular valuable for the box contributions, which are the most complex part of the calculation. First of all we found internal symmetries relating various form factors of one given box diagram to other form factors of the same diagram:
\begin{eqnarray}
  F_2 & = & F_3 \nonumber \\
F_6 & = & F_7\nonumber \\
F_9 & = & F_{10} \nonumber \\
F_1 & = &   4 \, m_e^2\, F_{11} + 2 \, m_e \,F_2       \nonumber \\
F_1 & = &   4 \, m_e \,F_{10} + 2 \, m_e \, F_2 + 2 \, m_e F_2       \nonumber \\
F_1 & = &   4 \, F_8 + 2 \, m_e \, F_2 \,\,\,\,   ( - 4 \, F_{12})  . 
\end{eqnarray}
The contribution in brackets is only present for so-called dotted-box diagrams, which will be presented in the next section.\\
Actually this symmetry somehow reflects the non-minimal choice of amplitude-basis we made. Further conventional crossing symmetry can be applied to relate $s$- and 
$t$- channel diagrams.
Moreover we found one additional symmetry relating  crossed box (cb) diagrams to direct box (db) diagrams in the same channel:
\begin{eqnarray}
F_1^{\rm cb}  & = & F_1^{\rm db} - 2\, m_e \, F_6^{\rm db} + 2 \, d\, F_8^{\rm db} - 4 \, m_e \, F_9^{\rm db} \nonumber \\
F_2^{\rm cb}  & = & - F_2^{\rm db} - 2\, d\, F_9^{\rm db} \nonumber \\
F_4^{\rm cb}  & = &  F_4^{\rm db} - (4-2\, d) \, F_{11}^{\rm db}\nonumber  \\
F_5^{\rm cb}& = &  - F_5^{\rm db} + (4 - 6\, d) F_{12}^{\rm db}\nonumber \\
F_6^{\rm cb} & = & -F_6^{\rm db} - 4\, F_9^{\rm db}\nonumber  \\
F_8^{\rm cb} & = & -F_8^{\rm db} \nonumber\\
F_9^{\rm cb} &=& F_9^{\rm db}\nonumber  \\
F_{11}^{\rm cb}& =& -F_{11}^{\rm db}  \nonumber  \\
F_{12}^{\rm cb} &= & F_{12}^{\rm db} ,  
\end{eqnarray}
and interchanging $F_{i}^{\rm cb}(s, t,u)  \leftrightarrow F_{i}^{\rm db}(s, u, t)    $ successively.
The space-time dimension  $d$ reflects our choice of dimensional regularization.
It is sufficient to calculate one box diagram and obtain the form factors
of the remaining three box diagrams via those symmetry relations. 

To reduce the various one-loop integrals we rely upon two different techniques. First we follow the prescription of Passarino and Veltman \cite{Passarino:1979jh} with the subtle difference from the original work that the tensor decomposition is performed with respect to inner momenta. 

Favourably we use the method first proposed by Davydychev  \cite{Davydychev:1991va} to reduce one-loop  tensor integrals into scalar integrals and changing the space-time dimension {\it d}, and later extended to multi-loop tensor integrals by Tarasov \cite{Tarasov:1996br}.  Let $  I_n^{\mu} (  I_n^{\mu\, \nu}) $ be a one-loop $n$-point vector (tensor) integral, $k$ the loop-momentum, $ c_r = (k-p_r)^2-m_r$  the propagator corresponding to the $r$-th line of the diagram appearing in the $\nu_r$-th power, with the special case $ c_n = k^2 - m_n^2$. Those integrals can be written as
\begin{eqnarray}
  I_n^{\mu} & =&  \int ^{d} k_{\mu} \prod_{r=1}^{n} \, \frac{1}{c_r^{\nu_r}}  - \sum_{i=1}^{n-1} \, p_i^{\mu} \, I_{n,i}^{[d+]} \\
 I_{n}^{\mu\, \nu}& =& \int ^{d} k_{\mu} \, k_{\nu} \, \prod_{r=1}^{n} \, \frac{1}{c_r^{\nu_r}} \\ \nonumber
&& \hspace{-1cm} = - \sum_{i,j=1}^{n-1} \, p_i^{\mu}\, p_j^{\nu} \, (1+ \delta_{i\, j}) \,  \, I_{n,ij}^{[d+]^2} -\frac{1}{2} \, g^{\mu \nu}  \, I_{n}^{[d+]} \, ,
\end{eqnarray}
where  $[d+]$ is an operator shifting the space-time dimension by two units and
\begin{eqnarray}
  \label{eq:Inij}
   I_{n, \, i\,j...} =  \int ^{d}  \prod_{r=1}^{n} \, \frac{1}{c_r^{\nu_r+\delta_{ri} + \delta_{rj}...}}
\end{eqnarray}
 is the original scalar integral with one additional power in the $i$-th and  $j$-th propagators. Having reduced the tensor integrals to scalar integrals the generic space-time dimension $ d $  needs to be re-established. For this we use the recurrence relations first proposed in \cite{Tarasov:1996br}, which are complementary to those obtained via integration by parts \cite{Tkachov:1981wb,Chetyrkin:1981qh}, and later 
  simplified and extended to zero Gram determinants in
\cite{Fleischer:1999hq}. 
Let $ Y_{ij}=-(p_i-p_j)^2+m_i^2+m_j^2$ and 
the Cayley determinant 
\begin{eqnarray}
()_n ~\equiv~  \left|
\begin{array}{ccccc}
  0 & 1       & 1       &\ldots & 1      \\
  1 & Y_{11}  & Y_{12}  &\ldots & Y_{1n} \\
  1 & Y_{12}  & Y_{22}  &\ldots & Y_{2n} \\
  \vdots  & \vdots  & \vdots  &\ddots & \vdots \\
  1 & Y_{1n}  & Y_{2n}  &\ldots & Y_{nn}
\end{array}
\right|,
\end{eqnarray}
then the so-called signed minors  $ {j_1\, j_2 ... \choose i_1 \, i_2 ...}_n $ are defined as the rows $j_1, j_2, ...$ and columns $ i_1, i_2, ...$ erased from the Cayley determinant $()_n$.\footnote{Note here the additional overall sign resulting from  $(-1) ^{j_1+j_2+...+i_1+i_2+..} $. } Making successive use of the following three
recurrence relations leads to scalar master integrals  $A_0, B_0, C_0 $ and  $D_0 $ in  $d $ dimensions:
{\small
\begin{eqnarray}
 \hspace{-3cm}\left(  \right)_n
 \nu_j{\bf j^+} I^{(d+2)}_n  & =& 
\left[  - {j \choose 0}_n +\sum_{k=1}^{n} {j \choose k}_n
 {\bf k^-} \right] I^{(d)}_n  \nonumber \\[2mm]
{0\choose 0}_n \nu_j {\bf j^+} I_n^{(d)}&\!  =& \!\!
 - \sum_{i,k \, i\neq k}^n   {0j\choose 0k}_n \,\, \nu_i \, {\bf
k^-} {\bf i^+}\, I_n^{(d)} +\nonumber \\ 
&& \hspace{-3cm}  \left[ \left( 1 + \sum_{i=1}^n \nu_i - d \right) {0\choose j}_n \! -
\sum_{k=1}^n  {0j\choose 0k}_n \! (\nu_k-1)   \right] \, I_n^{(d)} \nonumber \\[2mm]
&&  \hspace{-2.5cm} (d-\sum_{i=1}^{n}\nu_i+1) \left(  \right)_n  I^{(d+2)}_n  =  \nonumber \\
&& \hspace{-1cm}  \left[ {0 \choose 0}_n - \sum_{k=1}^n {0 \choose k}_n {\bf k^-} \right]I^{(d)}_n \, ,
\end{eqnarray}
} the operators ${\bf i^+, j^+} $ raising the power of the corresponding propagator by one unit , while ${\bf k^-} $ reduces the power of the  $k $-th propagator by one unit.
\noindent
For zero Gram determinants  $()_n=0 $,  similar recurrence relations are needed and are available in \cite{Fleischer:1999hq}. Effectively a zero Gram determinant reflects the kinematical boundaries of phase space where a given $n$-point function can be
expressed through scalar integrals of lower rank. Typical examples of such simplifications are
\begin{eqnarray}
C_0(m_e^2,s,m_e^2;0,m_e^2,m_e^2) &=& \nonumber \\
&& \hspace{-3cm}    - \frac{(d-2)}{(d-4)} \,
\frac{1}{s-4\,m_e^2} \, \frac{A_0(m_e^2)}{m_e^2} \nonumber \\
&& \hspace{-3cm} 
     + 2\, \frac{(d-3)}{(d-4)} \, \frac{1}{s-4\,m_e^2} \,
B_0(s;m_e^2,m_e^2) \nonumber \\[3mm]
&&  \hspace{-4cm}  B_0(m_e^2;m_e^2,0) = \frac{1}{2} \, \frac{ (d-2)}{(d-3)} \frac{A_0(m_e^2)}{m_e^2}   \, .
\end{eqnarray}
\section{Factorizable two-loop contributions}
Two-loop contributions to the Bhabha  matrix element can be classified according to the topological properties of the corresponding diagrams. In this section we focus on the contributions originating from the renormalization of the electron mass. To accommodate the renormalization contributions,  every internal electron propagator in Fig.~\ref{fig:1loop}  has to be replaced by
\begin{eqnarray}
  \label{eqn:massren}
\frac{1}{p_e^2-(m_e+\delta m_e)^2}\simeq\hspace{-1mm} \frac{1}{p_e^2-m_e^2}\hspace{-1mm}\left(\hspace{-1mm}
1+\frac{2 m_e \delta m_e}{p_e^2-m_e^2} \right)
\end{eqnarray}
with 
\begin{eqnarray}
   \delta m_e & =&  - e^2 \,  (1-\epsilon) \, \frac{A_0(m_e^2)}{m_e} \nonumber \\
&&      - 2 \, e^2 \, m_e \, B_0(p_e^2;m_e^2,0)  \, \nonumber
\end{eqnarray}
 the electron mass counter-term. From Eq.~(\ref{eqn:massren}) we understand that we have to calculate all possible permutations of the diagrams in Fig.~\ref{fig:1loop} with one of the loop-internal electron propagators squared. Now, since the  mass renormalization term $\delta_{m_e}$ contains 1/${\epsilon}$ UV poles, the 
 one-loop master integrals  $A_0, B_0, C_0 $ and  $D_0 $ are needed in  ${\it O}(\epsilon)$ to ensure that all finite terms are taken into account properly. The $\epsilon $-expansion of $A_0$ is pretty straightforward.
The hypergeometric presentation of 1-loop scalar master integrals in arbitrary dimensions $d $ was given in \cite{Tarasov:2000sf,Fleischer}. From these results the 
 $\epsilon $-expansion can be obtained and will be presented in \cite{FRTW:2002aa}.
The  electron propagator appearing squared does not really impose further complications in the recurrence relations approach. For this part of the calculation we refrain from further pursuing the independent calculation based upon the Passarino--Veltman reduction method.
\section{Results}
Given the scope of this article we cannot present the full analytical result here~\footnote{For a complete result see http://acat02.\linebreak[2]sinp\linebreak[2]\linebreak[2].msu.ru\linebreak[2]/presentations/\linebreak[2]fleischer/ACAT.ps}. To give an idea of the structure we present exclusively the form factors $F_5$ for the various $s$-channel topologies in the Appendix. We use the following abbreviations
\begin{eqnarray}
  \label{eq:abb}
A_0 & = & A_0(m_e^2) \,,  \nonumber \\
B_s &=& B_0(s;m_e^2,m_e^2)  \,, \hspace{0.5cm} B_t  =B_0(t;m_e^2,m_e^2) \,,  \nonumber \\
C_0 &=&C_0(m_e^2,m_e^2,s;0,m_e^2,0)  \,,\nonumber \\
D_0&=&D_0(m_e^2,m_e^2,m_e^2,m_e^2,t,s;m_e^2,0,m_e^2,0) \,, \nonumber \\
x^2&=&1/(1-4m_e^2/s) \,, \hspace{0.5cm}  y^2=1/(1-4m_e^2/t) \,, \nonumber \\
 z^2&=&\frac{4 m_e^2}{s}  \,, \hspace{0.5cm} r_t  = \frac{t}{s}  \,, \hspace{0.5cm} r_u = \frac{u}{s} \,.
\end{eqnarray}
\section{Summary}
In this article we have presented the complete one-loop originating contribution to the two-loop Bhabha cross section. We decomposed the matrix element into 12 amplitudes and presented the result in the form of the corresponding form factors. Mass renormalization lead to  dotted
diagrams, which we reduced to scalar integrals by making use of recurrence relations. 
Two-loop factorizable form factors were presented.
\section*{Acknowledgements}
\noindent
A.W. thanks the organizers for a very stimulating conference.
\addcontentsline{toc}{section}{References}

\def\thesection{\Alph{section}}
\def\theequation{\thesection.\arabic{equation}}
\setcounter{section}{0}
\setcounter{equation}{0}
\section{Appendix}
All form factors have the overall factor ${e^4}/{(4 \pi)^{d/2}}$. Form factors corresponding to dotted diagrams have the additional common factor ${ \delta_{m_e}}/{m_e}$. Both box contributions have the additional common factor $\rt^{-2} \ru^{-2}  $.  
\begin{eqnarray}
 F_5^{\rm self} &=&  \,  \left.  A_0 \frac{4} {s^2} [\frac{1}{d-1}-1] 
\bkr +\,   B_s \frac{2} {s} [\frac{1}{d-1} (1-z^2)-1] \right.\\ 
 F_5^{\rm dot \,self} & =& \,  \left.  A_0 \frac{4} {s^2} (d-2) x^2 z^2    \, \bkr - B_s \frac{2} {s  } z^2 [1-(d-3) x^2] \right. \\
 F_5^{\rm vertex} & =&  2\, \left(
-A_0 \frac{1}{m_e^2 s} [2 \frac{1}{d-3}-x^2 z^2]
\bkr +  \left( \frac{A_0}{m_e^2} + B_s \right) \frac{2}{(d-4)s} [1+x^2]\right) \bk  +\, B_s \frac{1}{s} [x^2 (1+z^2)+d-4]  \\
 F_5^{\rm dot\, vertex} & =&  2 \left(
 A_0 \frac{2} {s^2} x^2 \left[(\frac{1}{d-3}+6 \frac{1}{d-5}) \right. 
 \bkr  -(d-4) (2   x^2-1+(d-4))+7-4 x^2 \right]  \nonumber \\
&&  +\,A_0 \frac{2} {m_e^2 s} [1+3 \frac{1}{d-5}]
\bk  -\,B_s \frac{1} { s}  x^2 z^2 [(d-4) (2 x^2+(d-4))+2 x^2]  
\bk  \left.  - \left( \frac{A_0}{m_e^2} + B_s \right) \frac{2}{(d-4) s}  x^2 z^2 \right) 
\end{eqnarray}
\begin{eqnarray}
 F_5^{\rm dbox} &=&
 A_0 \frac{\rt}{ m_e^2  s}
 (-\frac{1}{(d-3)} (2 \rt (1-\rt \ru \bk -\rt^2+\rt \ru x^2)   +
                z^2 (\rt \ru-\rt+3 \rt^2-\ru y^2 \bk -z^2 \rt)    ) 
                -
               (2 \rt (1+\rt+2 \ru+\rt \ru x^2) \bk +2 z^2 (1+\ru) y^2)) 
 B_t \frac{4 \rt}{ s} (\rt (1+\rt+\ru) \bk +z^2 (1+\ru) y^2) -
B_s \frac{4 \rt}{ s} (\rt \ru (1+\rt x^2))                  
\bk  -
 \left( \frac{A_0}{{m_e}^2}+B_t \right) \frac{2 \rt}{ s  (d-4)} (2 \rt (\rt+\rt^2 \bk +(2+\rt) \ru)          
 +
               z^2 (\rt (1-\ru-3 \rt)
 \nonumber \\
 && \hspace{-1.3cm}+(2+3 \ru) y^2+z^2 \rt))-
C_0 \rt^2      (\frac{1}{(d-3)} \bk {\scriptscriptstyle \times} (2 (\rt-\rt^3+(1-\rt^2) \ru)-z^2 (2+\rt
+
               \ru \bk -3 \rt \ru-5 \rt^2
 -z^2 (1-\ru-4 \rt +z^2)))\bk  -
2 (\rt+\rt^2+(1+2 \rt) \ru-z^2 \rt (1-\ru x^2))) 
\bk 
 + 
 D_0 \rt^2 s    (\frac{1}{(d-3)} (\rt-\rt^3+z^2 (\rt+3 \rt^2-2 \rt^3  \bk   +
               2 (1-\rt^2) \ru-z^2 (2 (1+\rt)-5 \rt^2 +(1-3 \rt)
               \ru
              \bk-z^2 (1-\ru-4 \rt+z^2))                 )/2)+
               \rt (1+\rt) (1+\rt \bk +2 \ru)
              -z^2 (4 \rt (1+\rt)    \!  - \!
               ( \!\ru- \!4 \rt) \ru-2 z^2 \! \rt)/2) \nonumber \\
\end{eqnarray}
\begin{eqnarray}
F_5^{\rm dot \,dbox} &=& 
 A_0 \frac{1}{m_e^2 s} ((d-4) (2 \rt^2 (\rt^2+3 \rt+2 \bkd +\ru (2+2 \rt
+2 \ru-x^2 \rt))
-
                z^2 ( \rt (3 \rt^2+5 \rt \bkd +2 \rt \ru-y^2  
                (1+2 \ru) \ru)-z^2 (\rt (\rt-y^2) \bkd -(1+\ru) y^4) )) 
           +
              \frac{3}{(d-5)} (2 \rt^2 (\rt^2+3 \rt+2 \bkd +\ru (2+2 \ru+\rt)) -
                z^2 \rt (3 \rt^2+3 \rt+\rt \ru \bkd -y^2 2 (1+\ru+\ru^2)  -
                z^2 \rt))             
               +   
                2 \rt^2 (3 \rt^2+9 \rt \bkd +5 \rt \ru+6 (1+\ru+\ru^2)      -
                2 x^2 \rt \ru)+z^2 (\rt (1 \bkd +\ru-22 \rt-14 \rt^2
                -6 \rt \ru
                     +
                y^2 (3+17 \ru \bkd +12 \ru^2))/2                  +
                z^2 (\rt (8 \rt-2  -4 y^2 (1-\ru)) \bkd -4 y^4 (2+\ru))/4 ))
 -B_t \frac{z^2}{s} (2 (d-4) (2 \rt (\rt
\bkd +y^2 (1+\ru))+z^2 y^2 (\rt+y^2 (1+
                \ru)) )-\rt (1 \bkd +4 \rt^2+(1+4 \rt) \ru-5 y^2 (1-\ru))  +
         z^2 (\rt (1+ \bkd 2 \rt)      
-2 y^2 (\rt \ru-y^2))            )           
                -
 B_s     \frac{1}{s}  (d-3) (8 \rt^2 (1+\rt \bkd +\ru) \ru+z^2 \rt (1+3 \rt  +\ru 
                (1-4 \rt)+2 x^2 \rt (1+ \bkd 2 \rt) \ru
+2 y^2 (1+2 \ru+2 
                \ru^2)-z^2 (1+2 x^2 \rt \ru))  )                 \bkd  +\left( \frac{A_0}{{m_e}^2}+B_t \right) \frac{z^2}{(d-4) s} (\rt (1+\ru+4 \rt (1+\ru \bkd +\rt)
                -y^2 (1-9 \ru))
                -z^2 (\rt (1+2 \rt
 -2 y^2 (1+\bkd \ru))
 -2 y^4 \ru))       +
C_0      \rt      ((d-4) (4 \rt (1+\rt+\ru) \ru
\bkd +z^2 ((1  +\ru-\rt (1+10 \ru+
                8 \ru^2 +\rt (8+8 \ru+\bkd  4 \rt))  +4 x^2 \rt^2 \ru          +
                 2 y^2 (1+2 \ru+2 \ru^2))/2
                 -z^2 (1\bkd -4 \rt (1+  \ru)    - 6 \rt^2+y^2 (2+4 \ru+4 \ru^2) +2 z^2 \rt)\bkd /2))      -
                 z^2 (\rt  (\ru+2 \rt (1+\rt)-2 x^2 \rt \ru)          
 -
                 z^2 (1 \bkd+\ru +3 \rt+6 \rt^2 -z^2 (1+2 \rt))/2) )    
+D_0  \rt z^2 s   ((d\bkd -4) (\rt (1+2 (\rt+  (1+\rt+\ru) \ru)+\rt^2)        
-
                 z^2 (\rt\bkd (2+2 \ru+3 \rt)-y^2   (1+2 \ru+2 \ru^2)      -
   z^2 \rt)/2)\bkd
  -\rt (2+2 \rt+2 \rt \ru+  \ru^2 +3/2 \ru)    +
                 z^2 (\rt  (3+ \bkd 2 \ru)-y^2 (1+2 \ru+2 \ru^2))/2       )  
\end{eqnarray}

\end{document}